\documentclass[a4paper,12pt]{article}
\usepackage{hyperref} 
\pdfoutput=1
\usepackage{amsfonts,amsthm,amsmath,amssymb,graphicx,hyperref,youngtab,color,tcolorbox}
\numberwithin{equation}{section}
\usepackage{bbold}
\usepackage{bm}
\newtheorem*{theorem}{Theorem}
\begin{document}
\newtheorem{definition}{Definition}[section]
\newcommand{\be}{\begin{equation}}
\newcommand{\ee}{\end{equation}}
\newcommand{\bea}{\begin{eqnarray}}
\newcommand{\eea}{\end{eqnarray}}
\newcommand{\LE}{\left[}
\newcommand{\me}{\mathrm{e}}
\newcommand{\R}{\right]}
\newcommand{\nn}{\nonumber}
\newcommand{\Tr}{\text{Tr}}
\newcommand{\N}{\mathcal{N}}
\newcommand{\G}{\Gamma}
\newcommand{\vf}{\varphi}
\newcommand{\LL}{\mathcal{L}}
\newcommand{\Op}{\mathcal{O}}
\newcommand{\HH}{\mathcal{H}}
\newcommand{\arctanh}{\text{arctanh}}
\newcommand{\up}{\uparrow}
\newcommand{\down}{\downarrow}
\newcommand{\ket}[1]{\left| #1 \right>}
\newcommand{\bra}[1]{\left< #1 \right|}
\newcommand{\ketbra}[1]{\left|#1\right>\left<#1\right|}
\newcommand{\rd}{\partial}
\newcommand{\de}{\partial}
\newcommand{\ba}{\begin{eqnarray}}
\newcommand{\ea}{\end{eqnarray}}
\newcommand{\db}{\bar{\partial}}
\newcommand{\we}{\wedge}
\newcommand{\ca}{\mathcal}
\newcommand{\lr}{\leftrightarrow}
\newcommand{\f}{\frac}
\newcommand{\s}{\sqrt}
\newcommand{\vp}{\varphi}
\newcommand{\hvp}{\hat{\varphi}}
\newcommand{\tvp}{\tilde{\varphi}}
\newcommand{\tp}{\tilde{\phi}}
\newcommand{\ti}{\tilde}
\newcommand{\ap}{\alpha}
\newcommand{\pr}{\propto}
\newcommand{\mb}{\mathbf}
\newcommand{\ddd}{\cdot\cdot\cdot}
\newcommand{\no}{\nonumber \\}
\newcommand{\la}{\langle}
\newcommand{\lb}{\rangle}
\newcommand{\ep}{\epsilon}
 \def\we{\wedge}
 \def\lr{\leftrightarrow}
 \def\f {\frac}
 \def\ti{\tilde}
 \def\ap{\alpha}
 \def\pr{\propto}
 \def\mb{\mathbf}
 \def\ddd{\cdot\cdot\cdot}
 \def\no{\nonumber \\}
 \def\la{\langle}
 \def\lb{\rangle}
 \def\ep{\epsilon}

\title{\bf Tensor and Matrix models: a one-night stand or a lifetime romance?}
\author{\textsc{Pablo Diaz\thanks{pablodiazbe@gmail.com}}
\\
{\small \emph{$^{*}$Fields, Gravity \& Strings, CTPU,}} 
{\small \emph{Institute for Basic Science, Daejeon 34126 \rm KOREA}}}

\maketitle
\begin{abstract}
The spectra of energy eigenstates of free tensor and matrix models are organized by Kronecker coefficients and Littlewood-Richardson numbers, respectively. 
Exploiting recent results in combinatorics for Kronecker coefficients, we derive a formula that relates Kronecker coefficients with a hook shape with Littlewood-Richardson numbers. 
This formula has a natural translation into physics: the eigenstates of the hook sector of tensor models are in one-to-one correspondence with fluctuations of  1/2-BPS states in multi-matrix models. 
We then conjecture the duality between both sectors. Finally, we study the Hagedorn behaviour of tensor models with finite rank of the symmetry group and, using similar arguments, 
suggest that the second (high energy) phase could be entirely described by multi-matrix models.       
\\

\end{abstract}
\newpage
\tableofcontents

\section{Introduction}
There are several motivations for the study of tensor models in theoretical physics. From the Quantum Mechanics point of view, tensor models are expected to suit a description of entangled
 systems \cite{HW,CM}. From the quantum gravity perspective, and inspired by the success of matrix models in the description of two-dimensional quantum gravity \cite{earlierTM}, tensor models
 were proposed in the early 90's as a framework for studying higher dimensional quantum gravity \cite{random1,random2,random3}. Recently, the interest in tensor models has
 been boosted in the context of AdS$_2$/CFT$_1$ were the SYK model  \cite{SYK1,SYK2,SYK3,SYK4,SYK5,SYK6,SYK7,SYK8,SYK9,SYK10} has been shown to share
 the same large $N$ behaviour as a tensor model \cite{Witten}. Besides, the arrival of color tensor models \cite{color1,color2}, rainbow models \cite{MM3,MM4} and  multi-orientable models \cite{Ta1}, 
together with the understanding of their $1/N$ expansion
 \cite{1N1,1N2,1N3,1N4,Ta2,MM2} (which helped to resolve old large $N$ issues) has certainly triggered the rapid development of the subject in the last years.\\
In this work we plan to contribute to the development of tensor models and the physics they involve by putting them in contact with matrix models. In this regard, there are at least
 two hints which make us suspect that a connection (probably deep) between tensor and matrix models exists: 

\begin{enumerate}
\item The holographic conjecture of the SYK model in the tensor model version \cite{Witten} and the increasing suspicion that tensor models have holographic duals in broader contexts, seem to indicate that there should be an overlap between tensor and matrix models, since the latter have been proven to encode holographic duals at least in the large $N$ regime. Actually, finding the precise relation between both models will be extremely interesting for holography. On the one hand the dictionary between matrix models and gravity duals is well-understood nowadays but the perturbative expansion for multi-matrix models is highly complicated. On the other hand, tensor models with quartic interaction present an easy-to-tackle (melonic) perturbative expansion but its relation with gravity is still unknown. Therefore, understanding the connection between both theories will bring  insights into holography and perhaps new computational tools.  

\item The recent discovery of the relation between some sectors of  Kronecker coefficients and Littlewood-Richardson numbers in the field of combinatorics and group theory. This is relevant for us since Kronecker coefficients organize the spectrum of eigenstates of free tensor models \cite{1stpaper,Robert,GR2,2ndpaper}, whereas Littlewood-Richardson numbers have long been known to organize the spectrum of matrix models  \cite{CJR,restricted1,restricted2,restricted3,brauer1,brauer2,Brown:2007xh,Brown:2008ij}.   
\end{enumerate}

Based on (partially) matching the spectrum of the free theory in both models through the above-mentioned relation between Kronecker coefficients and 
Littlewood-Richardson numbers we conjecture in this paper that the corresponding tensor and matrix sectors are dual. Note that matching the spectrum is a 
necessary condition for duality, for a definite proof we should see that those sectors are dynamically equivalent, a programme that we leave for a future work.\\

The paper is organized as follows. We start by writing and evaluating the partition function with the singlet condition of the free tensor model in section \ref{pf}. 
The spectrum of energy eigenstates is organized by Kronecker coefficients, as can be read from eq. (\ref{squareKronecker}). 
This is in perfect agreement with the direct counting of invariants found in recent works. 
In section \ref{hookshape}, we gather some results on combinatorics of Kronecker coefficients and we derive (\ref{Kronhookformula}), 
which tells the explicit relation between the hook sector of Kronecker coefficients and Littlewood-Richardson numbers. 
We then manage to construct the matrix operators that those numbers count and we arrive at (\ref{matrixmatch}), 
which is the main result of the paper. Eq.  (\ref{matrixmatch}) tells us that the hook sector of the tensor model has the same 
number of eigenstates than a rather general multi-matrix set of operators. These multi-matrix operators are interpreted as encoding fluctuations 
 about a generic $\frac{1}{2}$-BPS state of a depth given by the length of the hook in the tensor sector.  In short, the hook sector of the 
tensor spectrum encodes the fluctuations of $\frac{1}{2}$-BPS states in the matrix theory. In section \ref{Hagedorn}, we return to the partition function. 
Tensor models with finite rank of the symmetry group are known to have Hagedorn behaviour, a fact that is interpreted as a phase transition at some finite 
temperature related to the rank of the group. We study the growth of states of the second (high energy) phase of tensor models and, using a known 
theorem of Kronecker coefficients, we conjecture that this second phase can be entirely described by a multi-matrix model. 
Finally, we include an appendix for the computation of Kronecker coefficients with a hook shape in the regime of large (but finite) rank.  
%%%%%%%%%%%%%%%%%%%%%%%%%%%%%%%%%%%%%%

\section{Tensor partition function} \label{pf}
Color tensors refer to tensors with no additional symmetry assumed. We will write a $d$-rank covariant color tensor as 
\begin{equation}\label{dtensor}
 \Phi=\Phi_{i_1i_2\dots i_d}~e^{i_1}\otimes \cdots \otimes e^{i_d},
\end{equation}
 where the complex-valued vectors $\{e^{i_k}, i_k=1,\dots, N_k\} $ form an orthonormal basis of the vector space $\mathbb{C}^{N_k}$. 
The components of the tensor transform under the action of \begin{equation}\label{gauge group}
 G_d\equiv U(N_1)\otimes \cdots \otimes U(N_d)
\end{equation}
as
\begin{equation}\label{unitaryaction}
 \Phi_{j_1j_2\dots j_d}=\sum_{i_1,\dots, i_d}U(N_1)_{j_1}^{i_1}\cdots U(N_d)_{j_d}^{i_d}  \Phi_{i_1i_2\dots i_d}.
\end{equation}
The complex conjugate of $\Phi$ is a $d$-rank contravariant tensor which transforms as
\begin{equation}\label{unitaryactionconjugate}
 \overline{\Phi}^{j_1j_2\dots j_d}=\sum_{i_1,\dots, i_d}\overline{U}(N_1)^{j_1}_{i_1}\cdots \overline{U}(N_d)^{j_d}_{i_d}  \overline{\Phi}^{i_1i_2\dots i_d}.
\end{equation}
The action of the free theory is simply
\begin{equation}
S=\Phi_{i_1i_2\dots i_d}\overline{\Phi}^{i_1i_2\dots i_d}.
\end{equation}
Invariant operators  are $n$-fold tensor products $\Phi^{\otimes n}\overline{\Phi}^{\otimes n}$, built out of $n$ copies of the tensor $\Phi$ contracted with $n$ copies of its conjugate. 
Each invariant is associated with the specific way indices of $\Phi$ and indices of $\overline{\Phi}$ are contracted subjected to a double coset equivalence, 
see \cite{GR} for details. Counting the number of invariant operators, building a basis which diagonalizes the two-point function of 
the free theory and computing correlators has been a recent subject of study \cite{1stpaper,MM1,Robert,GR2,2ndpaper}. 
In those studies it was manifest the prominent role of Kronecker coefficients in organizing the spectrum of energy eigenstates.\\

Let us write and evaluate the partition function of the color tensor free theory.\\
 In general, for a compact Lie group $G$, the multi-particle partition function of the free theory with the singlet condition (so, only $G$-invariant states are taken into account) can be written as \cite{S} 
\begin{equation}\label{partition function}
Z_G(t)=\int_G \text{d}g ~\text{exp}\Big(\sum_{n=1}^{\infty}\frac{1}{n}f(t^n)\chi(g^n)\Big),
\end{equation}
where $\text{d}g$ is the Haar measure and $f(t)$ is the single particle partition function which, in the case of one bosonic field is simply $f(t)=t$, $t$ being the letter that refers to the field. The function $\chi(g)$ stands for the character of the representation in which the field transforms. \\
In our case, the group is $G_d=U(N_1)\times \cdots \times U(N_d)$. As said before, the gauge invariant operators transform under the fundamental-antifundamental representation of $G_d$, so 
\begin{equation}\label{characterg}
\chi(g)=\text{tr}(g)\text{tr}(g^+), \quad g\in G_d.
\end{equation}
Now, since $\chi(g)$ is independent of the choice of basis, we will choose, for every $U(N_k)$, a basis where the matrix is diagonal. We will make use the of the Weyl parametrization of $U(N_k)$ and use $u_k\in U(N_k)$ to be $u_k=\text{diag.}(e^{i\theta_{k1}},\dots,e^{i\theta_{kN_k}})$, with $0\leq \theta_{ki}\leq 2\pi$. With this parametrization we will write a group integral as 
\begin{equation}
\int_{G_d}\text{d}g~F(g)=\prod_{k=1}^d\frac{1}{(2\pi)^{N_k}N_k!}\int_0^{2\pi}\prod_{j=1}^{N_k}\text{d}\theta_{kj}\prod_{1\leq l,m\leq N_k}|e^{i\theta_{kl}}-e^{i\theta_{km}}|^2 F(u_k).
\end{equation}
We will use the following convenient notation for the eigenvalues $z_{kj}=e^{i\theta_{kj}}$ and also use the string of eigenvalues
\begin{equation}
z_k\equiv (z_{k1},\dots,z_{kN_k}).
\end{equation}
First, let us notice that with this parametrization and using general properties of the Kronecker product of matrices, the character (\ref{characterg}) of a general element $g\in G_g$ can be written in terms of symmetric functions of the eigenvalues as 
\begin{equation}
 \chi(g^n)=\text{tr}(g^n)\text{tr}((g^+)^n)=p_n(z_1 z_2\cdots z_d)p_n(z^{-1}_1z^{-1}_2\cdots z^{-1}_d)
\end{equation}
where $p_n$ are the power sums and the string $z_1 z_2\cdots z_d$ stands for a collection of $\prod_{k=1}^dN_k$ variables of the type $z_{1i_1}\cdots z_{di_d}$ with $1\leq i_k\leq N_k$. We will also use the notation
\begin{equation}
\Delta(z_k)=\prod_{1\leq i,j\leq N_k}(z_{ki}-z_{kj})
\end{equation}
for the Vandermonde determinant. Thus, the partition function of the free tensor model can be written as an integral over complex eigenvalues as
\begin{equation}\label{partfunc}
Z_{G_d}(t)=\prod_{k=1}^d\frac{1}{(2\pi)^{N_k}N_k!}\oint\prod_{i=1}^{N_k}\frac{\text{d}z_{ki}}{z_{ki}}\Delta(z_k)\Delta(z^{-1}_k)~\text{exp}\Big(\sum_{n=1}^{\infty}\frac{t^n}{n}p_n(z_1\cdots z_d)p_n(z^{-1}_1\cdots z^{-1}_d)\Big).
\end{equation}
By Taylor expansion and reordering terms, it is not hard to see that the exponential in (\ref{partfunc}) can be expressed as a sum over partitions, that is,
\begin{equation}
\text{exp}\Big(\sum_{n=1}^{\infty}\frac{t^n}{n}p_n(z_1\cdots z_d)p_n(z^{-1}_1\cdots z^{-1}_d)\Big)=\sum_{n=1}^{\infty}\sum_{\lambda\vdash n}\frac{1}{z_{\lambda}}t^np_{\lambda}(z_1\cdots z_d)p_{\lambda}(z^{-1}_1\cdots z^{-1}_d).
\end{equation}
Using the relation between power sums and Schur functions 
\begin{equation}
p_{\lambda}(z)=\sum_{\mu\vdash n}\chi_{\mu}(\lambda)s_{\mu}(z)
\end{equation}
we may write
\begin{eqnarray}\label{exponential}
&&\text{exp}\Big(\sum_{n=1}^{\infty}\frac{t^n}{n}p_n(z_1\cdots z_d)p_n(z^{-1}_1\cdots z^{-1}_d)\Big)\nonumber \\
&&=\sum_{n=1}^{\infty}\sum_{\lambda, \mu,\nu\vdash n}\frac{1}{z_{\lambda}}t^n\chi_{\mu}(\lambda)\chi_{\nu}(\lambda)s_{\mu}(z_1\cdots z_d)s_{\nu}(z^{-1}_1\cdots z^{-1}_d).
\end{eqnarray}
Schur functions of the variables $z_1\cdots z_d$ (remember that they are $N$ variables) can be written as \cite{M}
\begin{equation}\label{Schurplen}
s_{\mu}(z_1\cdots z_d)=\sum_{\mu_1,\dots,\mu_d\vdash |\mu|}g_{\mu,\mu_1,\dots,\mu_d}s_{\mu_1}(z_1)\cdots s_{\mu_d}(z_d).
\end{equation}
The point of writing the exponential, and so the partition function in terms of Schur functions this way is because we can apply straightforwardly the explicit inner product of Schur functions \cite{Dolan}
\begin{equation}\label{expproduct}
\delta_{\mu\nu}=\langle s_{\mu},s_{\nu}\rangle_{N_k}=\frac{1}{(2\pi i)^{N_k}N_k!}\oint\prod_{i=1}^{N_k}\frac{\text{d}z_{ki}}{z_{ki}}\Delta(z_k)\Delta(z^{-1}_k)s_{\mu}(z_k)s_{\nu}(z_k), \quad l(\mu),l(\nu)\leq N_k.
\end{equation}
It is important to stress that Schur functions $s_{\mu}(z)$ are identically 0 whenever $l(\mu)$ is greater than the number of variables, as indicated in  (\ref{expproduct}). This will restrict the sums over partitions in the following.
By making first the  substitution of the exponential (\ref{exponential}) with (\ref{Schurplen}) in the partition function (\ref{partfunc}), and then applying (\ref{expproduct}) to each pair of Schur functions we arrive at
\small{\begin{eqnarray}\label{squareKronecker}
Z_{G_d}(t)&=&\sum_{n=1}^{\infty}\sum_{\substack{\lambda, \mu,\nu\vdash n\\
\mu_1,\dots\mu_d\vdash n\\
\nu_1,\dots\nu_d\vdash n\\
l(\mu_k),l(\nu_k)\leq N_k}}\frac{1}{z_{\lambda}}t^n\chi_{\mu}(\lambda)\chi_{\nu}(\lambda)g_{\mu,\mu_1,\dots,\mu_d}g_{\nu,\nu_1,\dots,\nu_d}\delta_{\mu_1\nu_1}\cdots \delta_{\mu_d\nu_d}\nonumber \\
&=&\sum_{n=1}^{\infty}\sum_{\substack{ \mu,\nu\vdash n\\
\mu_1,\dots\mu_d\vdash n\\
\nu_1,\dots\nu_d\vdash n\\
l(\mu_k),l(\nu_k)\leq N_k}}t^ng_{\mu,\mu_1,\dots,\mu_d}g_{\nu,\nu_1,\dots,\nu_d}\delta_{\mu_1\nu_1}\cdots \delta_{\mu_d\nu_d}\delta_{\mu\nu}\nonumber \\
&=&\sum_{n=1}^{\infty}\sum_{\substack{ \mu\vdash n\\
\mu_1,\dots\mu_d\vdash n\\
l(\mu_k)\leq N_k}}t^ng_{\mu,\mu_1,\dots,\mu_d}^2 \nonumber \\
&=&\sum_{n=1}^{\infty}\displaystyle{\left(\sum_{\substack{ 
\mu_1,\dots\mu_d\vdash n\\
l(\mu_k)\leq N_k}}g_{\mu_1,\dots,\mu_d}^2\right)}t^n,
\end{eqnarray}}
from where we can read that the number of gauge invariants operators is actually counted by the square of the Kronecker coefficients, with the suitable restriction on the permitted partitions for finite $N$. This result is in perfect agreement with the recent direct counting of invariant in tensor models. In the last line of (\ref{squareKronecker}) we have used the property
\begin{equation}
\sum_{\mu\vdash n} g_{\mu,\mu_1,\dots,\mu_d}g_{\mu,\nu_1,\dots,\nu_d}=g_{\mu_1,\dots,\mu_d}g_{\nu_1,\dots,\nu_d}
\end{equation}
when the sum over partitions $\mu$ is not restricted, as can be easily checked by the definition of Kronecker coefficients and the orthogonality relations of characters. \\
The evaluation of the partition function has been carried out for general tensors with $d$ indices and $G_d$ group of symmetry. For simplicity and without loss of generality,  we will consider $d=3$ and $G_3=U(N)^{\otimes 3}$ in the rest of the paper.

\section{Kronecker coefficients with a hook shape and multi-matrix models} \label{hookshape}
Although we do not know any combinatorial formula for computing general Kronecker coefficients, there are some broad families for which we know. The most remarkable of them is perhaps the family of Kronecker coefficients with a hook shape. Started in \cite{Bl} and refined in \cite{Liu}, this program succeeds in giving $g_{\mu\nu\lambda}$ a combinatorial interpretation when one of the partitions, say $\mu$, is a hook shape. In this section we will use their results to built a compact formula of Kronecker coefficients with a hook shape in terms of Littlewood-Richardson numbers. This formula will allow us to make contact with multi-matrix models, a correspondence that we will show in detail.  \\
Let us consider a hook partition with $n-r$ columns and $r+1$ rows and denote it $\mu(r)$, so $\mu(r)=(n-r,1^{r})$.
For $r=0,\dots,n-1$ the diagram $\mu(r)$ runs over all possible hook shapes. In \cite{Liu} it was shown that the Kronecker coefficients $g_{\mu(r)\nu\lambda}$ can be expressed in terms of the standard inner products\footnote{The standard inner product for symmetric functions is defined as $\langle s_{\lambda},s_{\mu}\rangle=\delta_{\lambda\mu}$, where $s_{\lambda}$ and $s_{\mu}$ are Schur functions, see \cite{M}.} of Schur and skew Schur functions as
\begin{equation}\label{liu}
g_{\mu(r)\nu\lambda}+g_{\mu(r-1)\nu\lambda}=\sum_{\gamma\vdash r}\langle s_{\lambda},s_{\nu/\gamma}s_{\gamma'}\rangle,
\end{equation}
where $\gamma'$ is the $\gamma$-transposed diagram, obtained from $\gamma$ by interchanging rows and columns. We will take eq. (\ref{liu}) as the starting point of our analysis. \\
Using the properties of products of Schur and skew Schur functions
\begin{equation}
c^{\lambda}_{\mu\nu}=\langle s_\lambda,s_{\mu}s_{\nu}\rangle=
\langle s_{\lambda/\nu},s_{\mu}\rangle, \quad \lambda\vdash n,~~ \mu\vdash n-r, ~~\nu \vdash r,
\end{equation}
 where  $c^{\lambda}_{\mu\nu}$ are the Littlewood-Richardson numbers, it is only a bit of work to find the compact expression
\begin{equation}\label{Kronhookformula}
 \boxed{~~g_{\mu(r)\nu\lambda}=\sum_{k=0}^{r}(-1)^{r+k}\sum_{\substack{\gamma\vdash k\\
\rho\vdash n-k}} c^{\nu}_{\rho\gamma}c^{\lambda}_{\rho\gamma'},\quad \nu,\lambda \vdash n}
\end{equation}
where the only partition of 0 is, by definition, $\emptyset$, and $c^{\lambda}_{\mu,\emptyset}=\delta^{\lambda}_{\mu}$. Equation (\ref{Kronhookformula}) shows explicitly the relation between Littlewood-Richardson numbers and Kronecker coefficients with one hook shape and it is completely general for partitions $\nu$ and $\lambda$. \\
Equation (\ref{Kronhookformula}) is suggestive from the physical point of view since, as we are going to see, it relates the spectra of energy eigenstates in tensor and matrix theories. As read from eq. (\ref{squareKronecker}), the spectrum of free tensor models is organized by the Kronecker coefficients, they measure the degeneracy of states (invariant operators) with energy $n$ as
\begin{equation}
\text{card}\{\mathcal{O}_n^{G_3-\text{Inv}}\}=\sum_{\mu,\nu,\lambda\vdash n}g^{2}_{\mu\nu\lambda}.
\end{equation}
 Thus, $g^2_{\mu(r)\nu\lambda}$ has a natural interpretation as counting the hook-shaped sector of the tensor model. \\
On the other hand, Littlewood-Richardson numbers have been long known to relate to the spectrum of multi-matrix models 
\cite{CJR,restricted1,restricted2,restricted3,brauer1,brauer2,Brown:2007xh,Brown:2008ij}. Specifically, for the case of two different bosonic matrices $Z$ and $Y$
\begin{equation}
\text{card}\{\mathcal{O}_{(n,m)}^{U(N)-\text{Inv}}\}=\sum_{\substack{\mu\vdash n+m \\
\nu\vdash n,~\lambda\vdash m}}(c^{\mu}_{\nu\lambda})^2,
\end{equation}
where $n$ and $m$  are the number of fields $Z$ and $Y$, respectively, which build the operators. One of the orthogonal basis of operators that relates to this counting is the restricted Schur basis. 
We will use it from now on. Restricted Schur operators
\begin{equation}\label{rS}
\chi_{(\mu;\nu,\lambda)ij}(Z,Y),\quad \mu\vdash n+m,~\nu\vdash n,~\lambda\vdash m, ~~i,j=1,\dots, c^{\mu}_{\nu\lambda}
\end{equation}
furnish a basis built on $n$ copies of $Z$ and $m$ copies of $Y$. See \cite{restricted2,restricted3} for details.\\
Now the question is: in terms of operators (\ref{rS}), what is the RHS of eq. (\ref{Kronhookformula}) counting?\\
First, realize that since $c^{\nu}_{\rho\gamma}$ counts the number of operators $\chi_{(\nu;\rho,\gamma)ii}(Z,Y)$, it is clear that
\begin{eqnarray}
\sum_{\substack{\gamma\vdash k\\
\rho\vdash n-k}} c^{\nu}_{\rho\gamma}c^{\lambda}_{\rho\gamma'}&=&\text{card}\Big\{\chi_{(\nu;\rho,\gamma)ii}(Z,Y)\chi_{(\lambda;\rho,\gamma')jj}(Z,Y)\Big|~\gamma\vdash k,~\rho\vdash n-k,\nonumber \\
&&i=1,\dots, c^{\nu}_{\rho\gamma},~~j=1,\dots, c^{\lambda}_{\rho\gamma'}\Big\}.
\end{eqnarray} 
  In order to take care of the alternating sum that appears in the RHS of  eq. (\ref{Kronhookformula}) we must restrict the set of multi-matrix operators under consideration. Let us first introduce some notation. We will call $\lambda\cap\mu$ the diagram formed from the common boxes of $\mu$ and $\lambda$ as we overlap them. The size of the intersection is always $|\lambda\cap\mu|\leq n$, saturating the inequality when $\lambda=\mu$. A partition is written as $\mu=(\mu_1,\dots, \mu_{l(\mu)})$. So, in the language of Young diagrams $\mu_i$ is the number of boxes of row $i$, and $l(\mu)$ is the number of rows of the diagram $\mu$. For instance, $\mu_{l(\mu)}$ is the number of boxes of the last row of diagram $\mu$.\\
It turns out that the alternating sum in the RHS of eq. (\ref{Kronhookformula}) is achieved by restricting $\rho$ to partitions whose last row has the same number of boxes as the last row of $\nu\cap\lambda$, that is,
\begin{equation}\label{comb}
g_{\mu(r)\nu\lambda}=\sum_{k=0}^{r}(-1)^{r+k}\sum_{\substack{\gamma\vdash k\\
\rho\vdash n-k}} c^{\nu}_{\rho\gamma}c^{\lambda}_{\rho\gamma'}=\sum_{\substack{\gamma\vdash r\\
\rho\vdash n-r\\
\rho_{l(\rho)}=(\nu\cap\lambda)_{l(\nu\cap\lambda)}}} c^{\nu}_{\rho\gamma}c^{\lambda}_{\rho\gamma'},\quad \nu,\lambda \vdash n.
\end{equation}
Actually, the alternating sum would have also been reproduced if we fix any other corner of $\rho$. The choice of the last row is convenient since the last box of the last row of any Young diagram is always a corner\footnote{Eq. (\ref{comb}) is a sophistication of the identity among combinatorial numbers
\begin{equation}
\binom{n-1}{m}=\sum_{k=0}^{m}(-1)^{m+k}\binom{n}{k}.
\end{equation}}.
With this observation we can write
\begin{eqnarray}
g_{\mu(r)\nu\lambda}&=&\text{card}\Big\{\chi_{(\nu;\rho,\gamma)ii}(Z,Y)\chi_{(\lambda;\rho,\gamma')jj}(Z,Y)\Big|~\gamma\vdash r,~\rho\vdash n-r,\nonumber \\
&&\rho_{l(\rho)}=(\nu\cap\lambda)_{l(\nu\cap\lambda)},~i=1,\dots, c^{\nu}_{\rho\gamma},~~j=1,\dots, c^{\lambda}_{\rho\gamma'}\Big\}.
\end{eqnarray} 
Now, for the sum of squares we have 
\begin{equation}\label{matrixmatch}
\boxed{
\begin{aligned}
&\sum_{\mu,\lambda\vdash n}g^2_{\mu(r)\nu\lambda}=\text{card}\Big\{\chi_{(\nu;\rho,\gamma)ii}(Z,Y)\chi_{(\lambda;\rho,\gamma')jj}(Z,Y)\chi_{(\lambda;\bar{\rho},\bar{\gamma}')kk}(Z,Y)\chi_{(\lambda;\bar{\rho},\bar{\gamma}')ll}(Z,Y)\Big|\\
&\gamma,\bar{\gamma}\vdash r~~\rho,\bar{\rho}\vdash n-r~~\nu,\lambda\vdash n~~
\rho_{l(\rho)}=(\nu\cap\lambda)_{l(\nu\cap\lambda)}~~i,j=1,\dots, c^{\nu}_{\rho\gamma}~~k,l=1,\dots, c^{\lambda}_{\bar{\rho}\bar{\gamma}'}\Big\}.
\end{aligned} }
\end{equation} 
Equation (\ref{matrixmatch}) tells us that the energy spectrum of the hook shape sector labeled by $\mu(r)$ of a 1-boson tensor model is in one-to-one correspondence 
with the set of rather general composite operators of multi-matrix models shown above.  The label $r$ of the hook tells the number of $Y$ fields which enter the operators in the matrix models. 
For $r=0$, there are only $Z$ fields ($n$ of them) and $\nu=\lambda$, so the matrix composites are (a power of) Schur polynomials $\chi_{\mu}(Z)$. 
Now, since the operators $\chi_{\mu}(Z)$ play an important role in $\mathcal{N}=4$ SYM for furnishing the $\frac{1}{2}$-BPS sector of the theory \cite{CJR}, 
and the product of Schur polynomials is a Schur polynomial, we will take license here and call  $(\chi_{\mu}(Z))^4$ $\frac{1}{2}$-BPS operators from now on.  \\

What we have done so far is to match the number of eigenstates of both theories in a certain sector. The match is highly non-trivial. 
Although this does not prove the duality between both theories since we should also match the dynamics of the fields, 
it clearly tells us that both theories are intimately related. Actually, we will find eq. (\ref{matrixmatch}) useful in order to interpret the hook sector of the 
tensor model given that in the context of matrix models restricted Schur operators have a well known meaning\footnote{Especially in the 
displaced corner approximation \cite{applications1,applications2,applications3,applications4,applications5}, where 
restricted Schur operators have been proven to be holographically dual to giant gravitons with strings attached.}.    \\
Now, the point here is to understand the meaning of the composite operators that appear in eq. (\ref{matrixmatch}). For that let us consider  the pieces
\begin{equation}\label{composites}
\mathcal{O}^{\nu\lambda}_{(\rho,\gamma)ij}=\chi_{(\nu;\rho,\gamma)ii}(Z,Y)\chi_{(\lambda;\rho,\gamma')jj}(Z,Y)\quad \gamma\vdash r,~\rho\vdash n-r,~\nu,\lambda\vdash n.
\end{equation}
First, realize that the operators $\mathcal{O}^{\nu\lambda}_{(\rho,\gamma)ij}$ are 0 if partitions $\nu$ and $\lambda$ differ in more than $r$ boxes, that is, if $|\nu\cap\lambda|<n-r$. As said above, $r=0$ forces $\lambda=\nu$. Let us think of the operators $\mathcal{O}^{\nu\nu}=\chi_{\nu}(Z)\chi_{\nu}(Z)$ for $r=0$ as the initial (unperturbed) states and consider the operators with $r=1,2,\dots$ as fluctuations of those states with increasing energy. We will interpret the parameter $r$ as the depth of the fluctuation. Thus, for $r=1$, for which $|\nu\cap\lambda|\geq n-1$, the operator\footnote{Note that for $r=0,1,2$ there are no multiplicities, so the latin indices are absent.} $\mathcal{O}^{\nu\lambda}_{(\rho,(1))}$ will encode a 1-box fluctuation of the state $\nu$ into $\lambda$, making explicit the transition state $\rho\vdash n-1$. The same applies for subsequent values of $r$, where the state $\nu$ turns into $\lambda$ after an $r$-box fluctuation.  Be aware that the process is symmetric, so the role of $\nu$ might have also been taken by $\lambda$. \\
In summary, the operators  (\ref{composites}) (and so (\ref{matrixmatch})) seem to give a complete description of the possible $Y$-driven fluctuations of $\frac{1}{2}$-BPS states in the matrix theory.

\section{Hagedorn phase transition for finite $N$}\label{Hagedorn}
Recently, It has been noticed that tensor models (of any classical gauge group) present such a rapid growth of states that there is no Hagedron behaviour in the limit $N\to \infty$ \cite{BAT,BKMT}. Actually, the partition function (\ref{squareKronecker}) is not even convergent for any finite value of $t$. This is because, as noticed in \cite{HW,GR}, 
\begin{equation} 
X_{\infty}(n)\equiv\sum_{\mu,\lambda,\nu\vdash n}g^2_{\mu\nu\lambda}=\sum_{\lambda\vdash n}z_{\lambda},
\end{equation}
and we know that
\begin{equation}\label{n!ap}
\lim_{n\to \infty}\frac{\sum_{\lambda\vdash n}z_{\lambda}}{n!}=1.
\end{equation}
Actually the convergence of (\ref{n!ap}) is quite fast. The reason for it is that the sum is dominated by the term associated with the one column Young diagram $z_{(1^n)}=n!$, the rest of the terms are subleading. Now, since $t$ is physically related to the temperature through $t=e^{-1/k_BT}$, the zero-radius convergence of the partition function series for $N\to \infty$ can be understood as the Hagedorn temperature going to 0 at that limit.\\
However, for finite $N$ the spectrum gets truncated since no states for which $l(\mu_k)>N$ are allowed. Actually, for finite $N$ the number of states is given by
\begin{equation}
X_N(n)\equiv\sum_{\substack{ 
\mu_1,\mu_2,\mu_3\vdash n\\
l(\mu_k)\leq N}}g_{\mu_1\mu_2\mu_3}^2,
\end{equation}
as can be read from (\ref{squareKronecker}).

The growth of $X_N(n)$ is then exponential at large $n$ and the system is expected to present Hagedorn behaviour with a temperature 
\begin{equation}\label{HT}
T_H (N)\sim \frac{1}{\log N},
\end{equation}
as noticed in \cite{BAT}. As usual, Hagedorn behaviour indicates the existence of a phase transition at $T_H$. So, if we start with a low energy state and we pump energy into the system the second phase will appear  at some point. The two phases will coexist from then on and the temperature will asymptotically stabilize at $T_H$. The partition function, which below $T_H$ is summable, describes one of the two phases. In this section we will argue that the phase that arises at high energy (whose states are not accounted in $Z_{G_3}(t)$) can be interpreted as a fluctuating $\frac{1}{2}$-BPS state, in a similar fashion as we treated the hook sector of the tensor model in the former section.\\
To support this claim we will give evidence that the number of invariant $n$-fold operators that are ``left out'' in $Z_{G_3}(t)$ for finite $N$ when $N<n$, namely $X_{\infty}(n)-X_N(n)$, possibly match the number of fluctuations of $\frac{1}{2}$-BPS states, when the energy of the fluctuations (depth) is given by  $n'=n-N-1$. 
We say ``possibly match'' since we will not be able to compute $X_N(n)$ exactly for $n'\geq 2$.

 \paragraph{Unperturbed $\frac{1}{2}$-BPS state.} Let us consider $n=N+1$ first, so $n'=0$, which is the energy at which the second phase is expected to appear. The number of states which are ``left out'' can be calculated exactly in this case. Notice, that these states must be labeled by three partitions where one should be the one-column, let us take it to be $\mu$, so $\mu=(1^n)$. In this case the Kronecker coefficients are easily calculated from the orthogonality properties of characters as
\begin{equation}
g_{(1^n)\nu\lambda}=\frac{1}{n!}\sum_{\sigma\in S_n}\chi_{(1^n)}(\sigma)\chi_{\nu}(\sigma)\chi_{\lambda}(\sigma)=
\frac{1}{n!}\sum_{\sigma\in S_n}\chi_{\nu'}(\sigma)\chi_{\lambda}(\sigma)=\delta_{\nu'\lambda},
\end{equation}
where $\nu'$ is the transposed diagram of $\nu$. Now, the number of states which are left out is
\begin{equation}\label{brane0}
X_{\infty}(N+1)-X_N(N+1)=\sum_{
\nu,\lambda\vdash N+1}g_{(1^{N+1})\nu\lambda}^2=\sum_{
\nu,\lambda\vdash N+1}\delta_{\nu'\lambda}=P_{N+1}.
\end{equation}
So, at the threshold energy $n=N+1$, the tensor model in its second phase presents the degeneracy of  (unexcited) $\frac{1}{2}$-BPS states labeled by $\nu\vdash n$. \\

\paragraph{Single fluctuation of  the $\frac{1}{2}$-BPS state.}
If we keep on pumping energy into the system, part of it will go into exciting modes labeled by partitions whose number of rows do not exceed $N$, and part of the energy will go to the second phase, exciting the $\frac{1}{2}$-BPS state. Let us take $n=N+2$, so $n'=1$. The states which are associated to the second phase must have $\mu=(1^{N+2})$ or $\mu=(2,1^{N})$. For the first option the counting follows the same path as before leading to a total of $P_{N+2}$ states. The number of states that correspond to  $\mu=(2,1^{N})$ can be calculated exactly, since the partition $(2,1^{N})$ is a hook\footnote{The partition $\mu=(2,1^{N})$  corresponds to $r=N$ with the convention we are using. Now, since $g_{\mu\nu\lambda}=g_{\mu'\nu'\lambda}$, as can be checked from the definition, $g_{\mu(N)\nu\lambda}=g_{\mu(1)\nu'\lambda}$.}. They match the number of operators $\mathcal{O}^{\nu'\lambda}_{(\rho,(1))}$ which, as seen before, are interpreted as 1-box fluctuations of the $\frac{1}{2}$-BPS state labeled by $\nu'$.

\paragraph{Small fluctuations of  the $\frac{1}{2}$-BPS state}
For $n'\geq 2$ the relevant states of the second phase are labeled by a partition $\mu$, made of one column of $N+1$ boxes and a diagram $\alpha$ with $n'$ 
boxes attached to the column (as shown in the figure), along with diagrams $\nu',\lambda\vdash n$. Note that for $n'\geq 2$ the 
Kronecker coefficients get harder to compute exactly.   For instance, if $n'=2$ we do not have yet a combinatorial method to  compute 
Kronecker coefficients of the type $g_{(2^2,1^{N-1})\nu'\lambda}$. In order to estimate those Kronecker coefficients we will use the following result:
\begin{theorem}[\cite{klem,dvir,cm}]
Let $\mu,\nu,\lambda\vdash n$, and denote $n'=n-\mu_1$, where  $\mu_1$ refers to the first row of $\mu$. Now,  if $n'<n-|\nu\cap\lambda|$  then $g_{\mu\nu\lambda}=0$.
\end{theorem}
This statement naturally  applies to the cases we are considering in this section, with $\mu$ as in the figure, so $n'=|\alpha|$. Actually, although  the result uses the first row of the diagram $\mu$ we can translate it into the first column of $\mu$ by changing $\mu\to\mu'$ and $\nu\to\nu'$, since $g_{\mu\nu\lambda}=g_{\mu'\nu'\lambda}$.
 The theorem clearly holds for hook shapes, but the usefulness of it relies on its application for general shapes $\mu$. In particular, for the partitions $\mu$ we are considering, 
the theorem reinforces the interpretation of $n'=|\alpha|$ being the depth of the fluctuation.   So, as for hook shapes, we will interpret the tensor 
states counted by $g_{\mu\nu'\lambda}$ with $\mu$ as in the figure as  $n'$-depth fluctuations of $\frac{1}{2}$-BPS states in matrix models labeled by $\nu'$. 

\begin{figure}[h]
\centering
\includegraphics[width=1.1\textwidth]{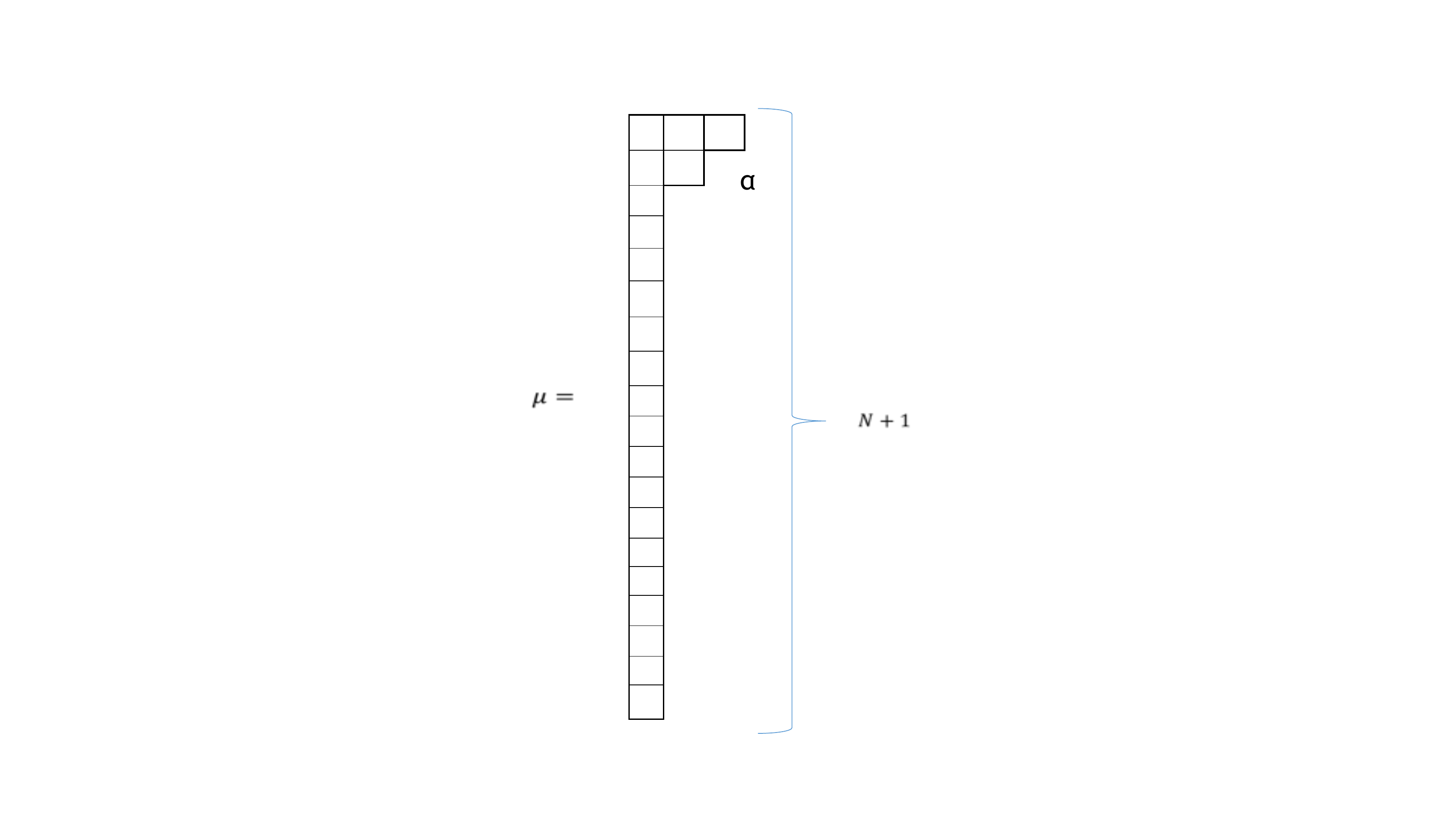}
\caption{Typical Young diagram for states of the second phase. Here $|\alpha|=n'$.}
\end{figure}\label{pict}

It is likely that there exist combinatorial formulas for $g_{\mu\nu\lambda}$ similar to  (\ref{Kronhookformula}), a result that would be extremely interesting in this discussion. Probably, the constraint  $\rho_{l(\rho)}=(\nu\cap\lambda)_{l(\nu\cap\lambda)}$ that appears in (\ref{matrixmatch}) (which is related to a 1-row diagram $\alpha$) should be replaced by any other related to a more general shape of $\mu$. We conjecture that this is actually the case and so the fluctuations are described by operators (\ref{composites}) with specific constraints for $\rho$ and  $\nu\cap\lambda$ related to $\mu$. This way the second phase of the tensor model would be described by a multi-matrix model.

To finish this section we will offer an estimation of the total number of states of the second phase for large $N$ and $n'\ll N$. We will use the conjecture above-mentioned and the corner approximations obtained in the appendix.  
The total number of states, calculated from (\ref{n'}) and (\ref{A}), is 
\begin{equation}
X_{\infty}(N+n'+1)-X_N(N+n'+1)\sim a_{n'} P_{N}N^{n'},
\end{equation}
where $a_{n'}\sim 2n'+1$ seems to hold\footnote{Numerical computations upto $n=100$ and $n'=7$ show a good agreement with these values of $a_{n'}$.}.

\section{Summary and outlook}
We have started the paper by writing and evaluating the partition function of free color tensor models with a symmetry group $G_d$. The partition function
shows that the spectrum of $G_d$-invariant energy eigenstates is organized by Kronecker coefficients, an expected result which serves as a consistency check.
Then, using recent mathematical algorithms for computing Kronecker coefficients with a hook shape we have derived eq. (\ref{Kronhookformula}). This identity has not appear in the 
literature before as far as we know and shows that, 
in the hook sector, Kronecker coefficients are computed by Littlewood-Richardson numbers. Now, the Littlewood-Richardson numbers are known to organize the spectrum of multi-matrix models. So, we step forward and interpret (\ref{Kronhookformula}) as relating both spectra of the theories in certain sectors. A precise match of the multi-matrix sector and a tensor hook shaped sector is shown in eq.  (\ref{matrixmatch}), which is the main result of the paper. Eq. (\ref{matrixmatch}) shows that for certain energy level determined by $n$, the different tensor states with a hook diagram $\mu(r)$ can be matched one-to-one with fluctuations driven by $r$ $Y$ fields of a Schur polynomial $\chi_{\mu}(Z)$ in the multi-matrix model. This strongly suggests that the 1-bosonic tensor model contains a multi-matrix model with two different species $Z$ and $Y$, each tensor state encoding a fluctuation of the Schur polynomial state.  The results so far apply to the hook sector of the tensor model. \\
One can see from the form of the partition function (\ref{squareKronecker}) that the partition function is not summable for $N\to \infty$, but it grows exponentially for finite $N$. This growth is a sign of a Hagedorn phase transition. Using known results for Kronecker coefficients we conjecture that the high energy Hagedorn phase, which appears at $n=N+n'+1$ for $n'=0,1,\dots$ could be  described by a multi-matrix theory and interpreted again as fluctuations of depth $n'$ of  Schur polynomial states.    \\

Given the match between spectra of both theories in the hook sector, a natural question is: Can we find a dynamical equivalence of both theories in the hook sector? This will be especially interesting for the interacting theory. Remember that the $SU(2)$ sector of $\mathcal{N}=4$ SYM is described by operators built on two matrix species $Z$ and $Y$, and that it is hard to tackle perturbatively. However, perturbative tensor models with a quartic interaction are known to lead to melonic Feynman diagrams which are much easier to handle.\\

It will be interesting to investigate the conjecture that the second phase, which appears at energies $n=N+n'+1$ for $n'=0,1,\dots$, is described by a multi-matrix theory with two species $Z$ and $Y$. We have conjectured that tensor states of the second phase correspond to fluctuations of Schur polynomial with depth $n'$. So the idea is to find the constraint on fluctuations for general shapes $\alpha$ analogous to $\rho_{\l(\rho)}=(\nu\cap\lambda)_{l(\nu\cap\lambda)}$ that appears in (\ref{matrixmatch}) for hook shapes. This is surely a tough problem, since finding the general rule would shed light on how to compute Kronecker coefficients using combinatorics, a mathematical problem which is lacking for a solution since 80 years ago.    \\

More generally, it will be desirable to investigate the relation between tensor and matrix models at the level of their actions. For instance, could matrix models appear as tensor model effective theories? This would clarify, for instance, if the flirt that we have shown in this paper is actually the beginning of a long term affair.

%%%%%%%%%%%%%%%%%%%%%%%%%%%%%%%%%%%%%%
\section*{Acknowledgement}
I thank S. Das, R. de Mello Koch, S-J. Rey and M. Walton for
useful discussions. This work was partly supported  by the Natural
Sciences and Engineering Research Council of Canada, the University of Lethbridge, and
by IBS-R018-D2.

\appendix
\section{Kronecker coefficients with a hook shape:\\
 Corner approximation}

In order to make estimations of $g^2_{\mu(r)\nu\lambda}$ and the sums which count the number of states, we will find useful to use the number of corners of a diagram, $C(\rho)$, 
which measures the number of boxes that can be deleted from diagram $\rho$ and still lead to a valid Young diagram. In the language of partitions, 
the number of corners is the number of different parts in partition $\rho$.  Thus, in this appendix we will arrive at approximations of  (\ref{Kronhookformula}) 
for which we only use the number of corners of the diagrams. We will trust these approximation in the regime of large $N$ and $n'\ll N$, 
in which case they are expected to reproduce the leading order (in $1/N$) correctly. 
\paragraph{Case $r=1$.}
For the simplest non-trivial case,  $r=1$, we will have $\mu(1)=(n-1,1)$ which is the transpose of $\mu(n-2)=(2,1^{n-2})$.  From the definition of Kronecker coefficients 
we immediately see that $g_{\mu(1)\nu\lambda}=g_{\mu(n-2)\nu'\lambda}$. With a diagram $\mu=(2,1^{n-2})$  we can find an exact formula for the sum of the square of Kronecker coefficients. 
A formula which involves only corners of partitions. \\
First, let us take two separate cases depending on whether the other two diagrams $\nu'$ and $\lambda$ are equal or not. 
If  $\nu'\neq\lambda$, where $\nu'$ is the transposed diagram of $\nu$, formula (\ref{Kronhookformula}) tells us that 
the Kronecker coefficient will be one if diagram $\lambda$ is obtained by taking a corner box from $\nu'$ and return it somewhere else. 
Otherwise it is zero. We can count all the possible non-zero combinations in the following way. Let us pick a diagram $\rho\vdash n-1$. 
The statement that  $\rho$ is connected (in the branching graph) with $\nu'$ and $\lambda$ means that adding a box to $\rho$ at one of 
its internal corners produces $\nu'$ and adding a box at a different internal corner produces $\lambda$. Now, if $\nu'$ and $\lambda$ are 
connected then the Kronecker coefficient $g_{(2,1^{n-2})\nu'\lambda}$ is 1. Given two different diagrams $\nu',\lambda\vdash n$ 
there is a unique $\rho\vdash n-1$ such that $\rho=\nu'\cap\lambda$. So, all the connected combinations $(\nu',\lambda)$ are found 
if we consider all diagrams $\rho\vdash n-1$, and for each one all possible ways of attaching a pair of boxes, that is, 
\begin{equation}\label{onebox}
\sum_{\nu'\neq\lambda}g^2_{(2,1^{n-2})\nu'\lambda}=\sum_{\nu'\neq\lambda}g_{(2,1^{n-2})\nu'\lambda}=\sum_{\rho\vdash n-1}C(\rho)(C(\rho)+1),
\end{equation}
where $C(\rho)$ is the number of corners of diagram $\rho$. In the first equality of (\ref{onebox}) we have used the fact that $g_{(2,1^{n-2})\nu'\lambda}$ is either 0 or 1.\\
When $\nu'=\lambda$ we can read from (\ref{Kronhookformula}) that 
\begin{equation}\label{eqdiagram}
g_{(2,1^{n-2})\nu'\nu}=C(\nu)-1.
\end{equation}
Gluing (\ref{onebox}) and (\ref{eqdiagram}) we obtain
\begin{equation}\label{suma}
\sum_{\nu,\lambda\vdash n}g^2_{(2,1^{n-2})\nu'\lambda}=\sum_{\rho\vdash n-1}C(\rho)(C(\rho)+1)+\sum_{\rho\vdash n}(C(\rho)-1)^2.
\end{equation}
This is an exact formula.

\paragraph{Case $r=2$.}
In (\ref{Kronhookformula}) we can see that the computation of the Kronecker coefficients for $r=2$, or equivalently for $r=N$ (our case) involves, at the most, diagrams with two boxes for $\gamma$.  
It is known that the only values that the Littlewood-Richardson numbers can take when one of the diagrams has two boxes or less are 0 or 1. 
So in this case we should not worry about multiplicities either. However, for the case $r=2$ it will not be possible to find an exact formula in terms of corners as we have done for $r=1$. 
For an exact formula we would need more information about the diagrams than just corners, like the number of double corners,  
which corresponds to rows from which we could remove two boxes and still arrive at a valid Young diagram. 
Nevertheless, we can make an estimation of the  order based on the number of corners. \\
First, realize that in (\ref{Kronhookformula}) the highest power of corners will happen when $k=2$ and it will be 4. 
For $k=1$ we saw in the paragraph above that the highest power was 2. In general, the highest power of corners in the sum will appear for $k=r$ and it will be $2r$. 
Now, since for large $N$ the sum will be clearly dominated by terms which involve the highest power of corners, 
for $r=2$ we will consider only $k=2$ bellow, and so $\gamma\vdash 2$ in (\ref{Kronhookformula}).

 Now, for $\gamma\vdash 2$, the product $c^{\nu'}_{\rho \gamma}c^{\lambda}_{\rho \gamma'}$ will be 1 if deleting two boxes from $\nu'$ and gluing them at
internal corners results in $\lambda$, provided that if the boxes deleted are in the same row (column) of $\nu'$ they are not in the same  row (column) of $\lambda$.  
Otherwise the product will be 0. We will not consider the cases where the boxes are deleted or placed at the same row/column. 
This restriction will allow us to still obtain the leading order at large $N$ using only the corners of the diagrams in our estimations. 
The number of pairs $(\nu',\lambda)$ which are left out with this restriction are not many. They are actually negligible 
for large $N$ since for diagrams with a large number of corners the number of choices of deleting (and gluing) two boxes at different places is 
much higher than the number of choices at the same row. So contributions from deleting/placing boxes at the same row/column will be always subleading. \\
We will distinguish three cases depending on $|\nu'\cap\lambda|$ being $n,n-1$ or $n-2$.\\
If $|\nu'\cap\lambda|=n$ then $\nu'=\lambda$. In this case we have $g_{(3,1^{n-3})\lambda'\lambda}\sim2\binom{C(\lambda)}{2}$, so
\begin{equation}\label{SO1}
\sum_{\rho\vdash n}g^2_{(3,1^{n-3})\rho'\rho}\sim\sum_{\rho\vdash n}4\binom{C(\rho)}{2}^2.
\end{equation}
If $\nu'\cap\lambda=\rho\vdash n-1$, so the diagrams differ in one box, then $g_{(3,1^{n-3})\nu'\lambda}\sim 2C(\nu'\cap\lambda)$. So
\begin{equation}\label{SO2}
\sum_{\nu'\cap\lambda\vdash n-1}g^2_{(3,1^{n-3})\nu'\lambda}\sim\sum_{\rho\vdash n-1}4C(\rho)^2C(\rho)(C(\rho)+1).
\end{equation}
If $\nu'\cap\lambda=\rho\vdash n-2$, so the diagrams differ in two boxes, then $g_{(3,1^{n-3})\nu'\lambda}= 2$, where 2 comes from the sum over $\gamma\vdash 2$, and
\begin{equation}\label{SO3}
\sum_{\nu'\cap\lambda\vdash n-2}g^2_{(3,1^{n-3})\nu'\lambda}\sim\sum_{\rho\vdash n-2}4\binom{C_i(\rho)}{2}\binom{C_i(\rho)-2}{2}\sim\sum_{\rho\vdash n-2}4\binom{C(\rho)+1}{2}\binom{C(\rho)-1}{2}.
\end{equation}
Consistently, we will take into account the sums of the contributions coming from $C(\rho)^4$. From (\ref{SO1}), (\ref{SO2}) and (\ref{SO3}) we see that 
\begin{equation}\label{SO}
\sum_{\nu,\lambda\vdash n}g^2_{(3,1^{n-3})\nu\lambda}\sim 6\sum_{\rho\vdash n}C(\rho)^4,
\end{equation}
where we have taken into account that 
\begin{equation}
\frac{\sum_{\rho\vdash n-n'}C(\rho)^r}{\sum_{\rho\vdash n}C(\rho)^r}\longrightarrow 1, \quad n\to \infty.
\end{equation}

\paragraph{General $r=n'$.}
When $n=N+n'+1$ with $n'\ll N$ we will be interested in calculating sums of $g^2_{(n'+1,1^{N})\nu'\lambda}$. 
When we estimate the sums using corners we will have terms in the sum like $C(\rho)^{2n'}$ which will 
dominate the sum. So we will consider those terms only. Also, the approaches taken for the case $r=2$ will apply here. 
Be aware that all these approximations make sense only for large $N$ and $n'\ll N$. Notice that for the cases we consider in these approximations where $n'$ boxes are deleted from $\nu'$ at {\it different} corners, the Littlewood-Richardson numbers are $c^{\nu'}_{\rho\gamma}=d_{\gamma}$, where $d_{\gamma}$ is the dimension of the representation of the symmetric group labeled by partition $\gamma$. So  $c^{\nu'}_{\rho\gamma}c^{\nu'}_{\rho\gamma}=d^2_{\gamma}$. Now, the sum in $\gamma$ that appears in (\ref{Kronhookformula}) can be performed to give a factor $n'!$, since $\sum_{\gamma\vdash n'}d_{\gamma}^2=n'!$. \\
All in all, the total sum can be approximated (to leading term) as
\begin{equation}\label{n'}
\sum_{\nu',\lambda\vdash N+n'+1}g^2_{(n'+1,1^{N})\nu'\lambda}\sim A(n') \sum_{\rho\vdash n}C(\rho)^{2n'},
\end{equation} 
with
\begin{equation}\label{A}
A(n')=n'!^2\sum_{m=0}^{n'}\binom{n'}{m}^2\binom{n'}{n'-m}^2.
\end{equation}

\end{document}